\documentclass[twocolumn,showpacs,preprintnumbers,amsmath,amssymb]{revtex4} 
\input psfig.sty

\usepackage{graphicx}
\usepackage{dcolumn}
\usepackage{bm}

\begin{document}

\title{Current and future supernova constraints on decaying $\Lambda$ cosmologies}

\author{J. S. Alcaniz} \email{alcaniz@astro.washington.edu}

\affiliation{Astronomy Department, University of Washington,
Seattle, Washington, 98195-1580, USA}

\author{J. M. F. Maia} \email{jmaia@axpfep1.if.usp.br}
\affiliation{Conselho Nacional de Desenvolvimento Cient\'{\i}fico e Tecnol\'ogico, SEPN 509, BL A, Ed. Nazir
I, 4$^{\underline o}$ andar, COCEX 70750-901, Bras\'{\i}lia - DF, Brazil
}

\date{\today}


\begin{abstract}
We investigate observational constraints from present and future
supernova data on a large class of decaying vacuum cosmologies. In
such scenarios the present value of the vacuum energy density is
quantified by a positive $\beta$ parameter smaller than unity. By
assuming a Gaussian prior on the matter density parameter
($\Omega_{\rm{m}} = 0.35 \pm 0.07$) we find $\Omega_{\rm{m}} =
0.34^{+0.14}_{-0.12}$ and $\beta = 0.62^{+0.12}_{-0.24}$ ($95\%$
c.l.) as the best fit values for the present data. We show that,
while the current data cannot provide restrictive constraints on
the $\Omega_{\rm{m}} - \beta$ plane, the future SNe data will limit
considerably the allowed space of parameter. A brief discussion about the equivalence between
dynamical-$\Lambda$
scenarios and scalar field cosmologies is also included.
\end{abstract}

\pacs{98.80-k; 98.80.Es; 98.62.Ai; 95.35+d}
\maketitle
\section{Introduction}

A large number of astronomical observations have led to a
resurgence of interest in a Universe dominated by a relic
cosmological constant ($\Lambda$). The basic set of experiments
includes the luminosity distance measured from type Ia supernova
(SNe Ia) \cite{perlmutter,riess}, measurements of cosmic microwave
background (CMB) anisotropies \cite{bern}, clustering estimates
\cite{calb}, age estimates of globular clusters \cite{carreta} and
high-redshift age estimates \cite{dunlop}. It is believed that the
presence of an unclustered component like the vacuum energy  not
only explains the observed accelerated expansion but also
reconciles the inflationary flatness prediction
($\Omega_{\rm{Total}} = 1$) with the $\Omega_{\rm{m}}$ measurements
that point sistematically to a value between $\Omega_{\rm{m}} = 0.2
- 0.4$.

On the other hand, it is also well known that the same welcome
properties that make models with a relic cosmological constant
($\Lambda$CDM) our best description of the observed universe also
result in a serious fine tuning problem \cite{wein}. The basic
reason is the widespread belief that the early universe evolved
through a cascade of phase transitions, thereby yielding a vacuum
energy density which is presently 120 orders of magnitude smaller
than its value at the Planck time. Such a discrepancy between
theoretical expectations and empirical observations constitutes
what is usually called ``the cosmological constant problem", a
fundamental question at the interface uniting astrophysics,
cosmology and particle physics.

As a phenomenological attempt of alleviating such a problem, the
so-called dynamical-$\Lambda$ or decaying $\Lambda$ cosmologies
were originally proposed in \cite{ozer}. Afterwards, a number of
different scenarios with suggestive  decaying laws for the
variation of the cosmological term were investigated
\cite{ozer1} (see also \cite{over} for a review). In Refs.
\cite{jackson,lima1} the authors proposed a class of deflationary
cosmologies driven by a decaying vacuum energy density whose
present value, $\Lambda_o$, is a remnant of a primordial
deflationary stage. Such models are analytic examples of warm
inflation scenarios proposed more recently by Berera \cite{WI} in
which particle production occurs during the inflationary period
and, as consequence, the supercooling process, as well as the
subsequent reheating are no longer necessary. The basics of the
cosmological history of such models can be summarized as follows
\cite{jailson}: first, an unstable de Sitter configuration is
supported by the largest value of the vacuum energy
density. This nonsingular state evolves to a
quasi-Friedmann-Roberton-Walker (FRW) vacuum-radiation phase and,
subsequently, the Universe approaches continuously to the present
vacuum-dust stage. The first stage harmonizes the scenario with the
cosmological constant problem, while the transition to the second
stage solves the horizon and other well-know problems in a similar
manner as in inflation. Finally, the Universe enters in the present
accelerated vacuum-dust phase as apparently suggested by the SNe Ia
observations. Other specific examples of exact
deflationary models and the underlying thermodynamics have been
studied recently by Gunzig {\it et al.} \cite{GMN}. Some scalar
field motivated descriptions for this class of models were investigated in Refs.
\cite{maar,zimd,jackson1}.

From the observational viewpoint some analyses have indicated a
good agreement between these models and different classes of
comological tests \cite{joao}. For example, Bloomfield Torres and
Waga \cite{tw} analysed gravitational lensing constraints on a
scenario in which the cosmological term decreases with time as
$\Lambda\propto R(t)^{m}$, where $R(t)$ is the scale factor and $m$
is a free parameter in the interval $0 \leq m \leq 3$. They found
that for low values of $\Omega_{\rm{m}}$ there is a wide range of
values of $m$ for which the lensing rate is considerably smaller
than in the conventional $\Lambda$CDM models. In particular, for
values of $m \geq 1$, these models reproduce very well the observed
lens statistics in the Hubble Space Telescope Snapshot Survey. More
recently, Vishwakarma \cite{vish} investigated some $\Lambda(t)$
models in the light of SNe Ia data and measurements of the angular
size of compact radio source ($\theta(z)$). The same $\theta(z)$
data together with age constraints from globular clusters and
high-$z$ galaxies were also analysed by Cunha {\it et al.}
\cite{jailson} in the context of deflationary scenarios. In all of
these cases, a good agreement between theory and observations was
found.

The aim of the present paper is twofold: first, to investigate the
observational contraints from the current SNe Ia data on the class
of deflationary cosmologies proposed in \cite{jackson,lima1} and
compare them with other recent results; second, to simulate future
SNe Ia data to infer how restrictive will be the limits on the
decaying rate $\beta$ from these high quality data. By future SNe
Ia data we assume a large number of high-$z$ supernova that will
become available from the projected Supernova/Acceleration Probe
(SNAP) satelite mission \cite{snap}. Such data have been explored
in the recent literature by several authors aiming mainly at
determining a possible time or redshift dependence of the dark
energy \cite{al}. A scalar field description for our dynamical-$\Lambda$ scenarios is also discussed.

This paper is organized as follows. In the next section we present the basic field equations relevant for our
analysis. In Sec. III we discuss a procedure to find scalar field counterparts for the decaying-$\Lambda$
cosmologies
studied here which share the same dynamics and
temperature evolution law. The corresponding constraints for deflationary cosmologies from the current SNe Ia
data are investigated in Sec. IV. In Sec. V we discuss the improvements in the parameter estimation by
simulating the future SNAP data. We end this paper by summarizing the main results in the conclusion section.

\section{The model: basic equations}

For homogeneous and isotropic cosmologies driven by nonrelativistic matter plus a cosmological
$\Lambda(t)$-term the
Einstein field equations are given by
\begin{equation}
\label{rho} 8\pi G \rho + \Lambda(t) = 3 \frac{\dot{R}^2}{R^2} + 3
\frac{k}{R^2} \quad,
\end{equation}

\begin{equation}
\Lambda(t) = 2 \frac{\ddot{R}}{R} +
\frac{\dot{R}^2}{R^2} + \frac{k}{R^2}\quad ,
\end{equation}
where an overdot means time derivative, $R(t)$ and $k = 0, \pm 1$
are, respectively, the scale factor and the curvature parameter and
$\rho$ stands for the dust energy density.

In this kind of deflationary cosmologies, the effective $\Lambda
(t)$ term is a dynamic degree of freedom that relaxes to its
present value, $\Lambda_o$, according to the following ansatz
\cite{lima1}
\begin{equation} \label{ansatz}
\rho_{v} = \frac{\Lambda(t)}{8 \pi G} = \beta \rho_{T} \left(1 +
\frac{1 - \beta}{\beta} {H \over H_{I}}\right) \quad ,
\end{equation}
where $\rho_{v}$ is the vacuum density, $\rho_{T}=\rho_{v}+ \rho$
is the total energy density, $H={\dot{R}}/R$ is the Hubble
parameter, $H_{I}^{-1}$ is the arbitrary time scale characterizing
the deflationary period, and $\beta\in[0,1]$ is a dimensionless
parameter of order unity. As shown in \cite{jackson,jailson}, regardless
the choice for the begining of the deflationary process, the scale
$H_{I}$ is unimportant during the vacuum-dust dominated phase ($H
<< H_{I}$)  so that, for all practical purposes at the late stages
of the cosmic evolution, the vacuum energy density (Eq. 3) can be
approximated by $\rho_v = \beta \rho_{T}$. It is worth mentioning that only for flat scenarios $\beta$ can be
considered identical to the vacuum energy density $\Omega_\Lambda$. From Eqs. (1) and (3) it is possible to
show that the general relation between these two parameters is given by $\Omega_\Lambda = \beta[1 +
k/(RH)^{2}]$.

By combining the above equations, we see that the deceleration parameter,
usually defined as $q_o = -R\ddot{R}/\dot{R}^{2}|_{t_{o}}$, now
takes the following form
\begin{equation}
q_o  =  \frac{\Omega_{\rm{m}}}{2} \frac{(1 - 3\beta)}{(1 - \beta)}
\end{equation}
or still, in terms of the curvature parameter $k$, 
\begin{equation}
q_o = {1 - 3\beta \over 2} \left[ 1+ {k \over (R_oH_o)^2} \right],
\end{equation}
which clearly reduces to the standard relation $q_o  =
\frac{\Omega_{\rm{m}}}{2}$ in the limit $\beta \rightarrow 0$.
As can be seen from the above equation, for any value of
$\Omega_{\rm{m}} \neq 0$, the deceleration parameter with a
decaying vacuum energy is always smaller than its corresponding in
the standard context and the critical case, $\beta = 1/3$ ($q =
0$), describes exactly the class of ``coasting cosmologies" studied
in \cite{kolb}.

\section{Scalar Field Version}

         A procedure to find scalar field counterparts for flat
decaying-$\Lambda$ cosmologies sharing the same dynamics and
temperature evolution law  has been  recently proposed \cite{jackson1}.
In this Section, we extend such a procedure to write the necessary
equations of a ``coupled quintessence'' \cite{cquint} version for
the late time behavior of the arbitrary $k$ cosmology considered
here.

Firstly, we extend the analysis of Ref. \cite{jackson1} and use
Eqs. (1)-(3) in the limit $H/H_I \to 0$ (our present time universe)
to define the parameter
\begin{equation}\label{eq:gstar}
\gamma_* \equiv -{2\dot H \over 3H^2} = (1- \beta ) \left( 1 + {k\over (RH)^2}\right).
\end{equation}
The above equation is just another way of writting the field
equations, so that any cosmology dynamically equivalent to the
asymptotic decaying-$\Lambda$ model considered here must have the
same $\gamma_*$.  Perhaps this fact can be made more explicit if
one considers that such a parameter is directly related to the
deceleration parameter:
\begin{equation}
\gamma_* = {2\over 3} (q+1).
\end{equation}
From Eqs. (1)-(3) it can be found that
\begin{equation}\label{eq:1stint}
H^2 + {k\over R^2} = A R^{-3(1-\beta)},
\end{equation}
where the integration constant $A=(8\pi G /3) \rho_{T0}
R_{0}^{3(1-\beta)}$. Another useful parameter necessary to simplify the notation
of the scalar field equations is
\begin{equation}
x \equiv {\dot\phi^2 \over \dot\phi^2 + \rho - 2k / 3(RH)^2},
\end{equation}
which reduces to the $x$ parameter used in \cite{jackson1} for $k=0$. In the above equation, $\phi$ is a
minimally coupled scalar field and, as indicated earlier, an overdot denotes time derivative.
Now, if we define the scalar field energy density and pressure:
$$ \rho_{\phi} = {\dot\phi \over 2} + V(\phi ) \quad   \mbox{and} \quad    p_{\phi} = {\dot\phi \over 2} -
V(\phi ),$$
respectively, and replace in Eqs. (1) and (2) the vacuum energy density and
``pressure''
 $$\rho_v = \Lambda /8\pi G \to \rho_{\phi} \quad   \mbox{and} \quad    p_v = - \Lambda / 8\pi G \to
p_{\phi},$$ 
 by these scalar field counterparts, we can manipulate the resulting equations to obtain (see \cite{jackson1}
for more details)
\begin{equation}
\rho = {3H^2 \over 8\pi G}\gamma_* (1-x),
\end{equation}
\begin{equation}\label{eq:dotphi}
\dot\phi^2 =  {3H^2 \over 8\pi G} \gamma_* x,
\end{equation}
\begin{equation}\label{eq:V}
V(\phi) = {3H^2 \over 8\pi G} \left[1 - \gamma_* \left(1-{x\over
2}\right)\right],
\end{equation}
so that, using (\ref{eq:1stint}) and (\ref{eq:dotphi}) we can find
(except for an integration constant)
\begin{equation}
\phi =  \sqrt{3(1-\beta)\over 8\pi G } \int{{AR^{-3(1-\beta)+2} - {2\over3}k \over AR^{3(1-\beta)+2} -
k}x^{1\over 2}{dR\over
R}}.
\end{equation}
For a given parameter $x(R)$, we can integrate the above
expression and study the behavior of the scalar field
potential parametrically. The explicit funtional relation for $x$
does not depend on the dynamics, so that for all practical purposes
it does not alter the SNe analysis presented here and could be
arbitrarily chosen. However, simplifying assumptions allows a
direct solution for the integral. If one imposes that the scalar
field version mimics exactly the particle production rate of its
decaying-$\Lambda$ counterpart, it can be shown that $x={\rm
const.}$. If, additionally, we consider the flat case, as suggested
by recent CMB data, we find the simple potential (a similar
reasoning was used in Ref. \cite{jackson1})
\begin{equation}
V( \phi ) \propto e^{-\lambda\phi},
\end{equation}
where $\lambda = \sqrt{3\pi G(1-\beta) / 2x}$.

\begin{figure}
\vspace{.2in}
\centerline{\psfig{figure=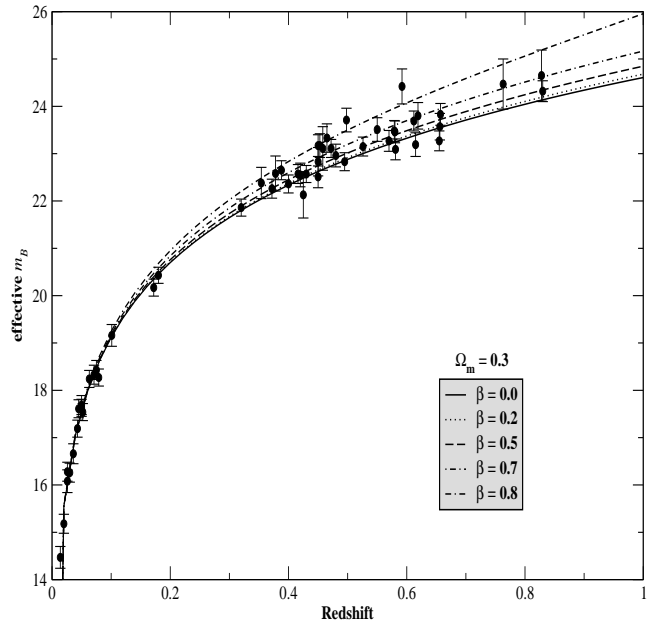,width=3.5truein,height=3.5truein,angle=-90}
\hskip 0.1in}
\caption{Hubble diagram for 16 low-redshift supernova and 38 high-redshift supernova from
\cite{perlmutter}.  As indicated in the Figure, the curves
correspond to a fixed value of $\Omega_{\rm{m}} = 0.3$ and several
values of $\beta$. For the sake of comparion the standard open
case, $\beta = 0$, is also shown (solid curve).}
\end{figure}

\section{Constraints from current SNe Ia data}

The apparent magnitude $m$ of a supernovae at a redshift $z$ is given by
\begin{equation}
m(z) = {\cal M} + 5\mbox{log}_{10}{\cal D}_{\mbox{{\scriptsize
L}}}(\Omega_{\rm{m}},\beta,z)
\end{equation}
where ${\cal M} = M - 5\mbox{log}_{10}H_0 +25$ is the ``zero point
magnitude", $M$ is the absolute magnitude of the supernovae and
${\cal D}_{\mbox{{\scriptsize L}}}(\Omega_{\rm{m}},\beta,z) \equiv
H_od _{\mbox{{\scriptsize L}}}(\Omega_{\rm{m}},\beta, H_o, z)$ is
the dimensionless luminosity distance written as
\begin{eqnarray}
{\cal D}_{\mbox{{\scriptsize L}}}(\Omega_{\rm{m}},\beta,z) & = & (1
+ z) \times
\\ \nonumber & &\int_{(1+z)^{-1}}^{1} {dx \over x\left[1-\frac{\Omega_{\rm{m}}}{1 - \beta} +
\frac{\Omega_{\rm{m}}}{1 - \beta} x^{-(1-3\beta)}\right]^{1/2}}.
\end{eqnarray}
The above expression can be integrated yielding
\begin{eqnarray}
{\cal D}_{\mbox{{\scriptsize L}}}(\Omega_{\rm{m}},\beta,z)& = & (1
+ z) \times
\\  \nonumber & & \frac{{\rm{sin}}\left[\delta
\rm{sin}^{-1}\Delta_1-\delta \rm{sin}^{-1}\Delta_2\right]}{\left(\frac{\Omega_{\rm{m}}}{1 - \beta} -
1\right)^{\frac{1}{2}}},
\end{eqnarray}
where $\delta=\frac{2}{(1-3\beta)}$, $\Delta_1=
(1-\frac{1-\beta}{\Omega_{\rm{m}}})^{{1 \over 2}}$ and $\Delta_{2}=\Delta_{1} (1 +
z)^{-(\frac{1-3\beta}{2})}$.
For $\frac{\Omega_{\rm{m}}}{1 - \beta} = 1$ or, equivalently, $k = 0$ (flat case) the above equation reduces
to
\begin{equation}
{\cal D}_{\mbox{{\scriptsize L}}}(\beta,z) = \frac{2}{1 - 3\beta}\left[(1 + z) - (1 + z)^{\frac{1 +
3\beta}{2}}\right].
\end{equation}

\begin{figure}
\vspace{.2in}
\centerline{\psfig{figure=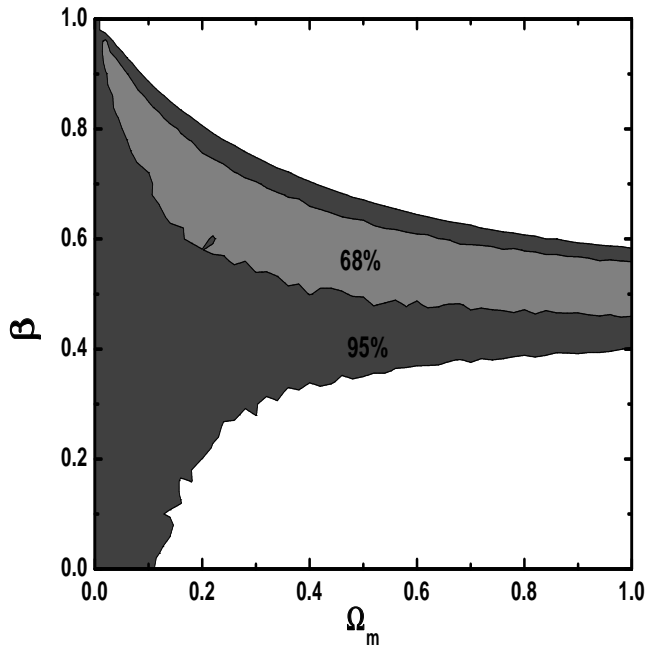,width=3.5truein,height=3.5truein}
\hskip 0.1in}
\caption{Confidence regions (68$\%$ and 95$\%$) in the $\Omega_{\rm{m}} - \beta$ plane provided by the
SNe Ia data from Perlmutter {\it et al.} \cite{perlmutter}.}
\end{figure}

In this analysis we consider the SNe Ia data set from the Supernova
Cosmology Project (SCP) \cite{perlmutter}. As noted in
\cite{perlmutter}, from the 60 supernova events, some are
considered outliers so that we work with a total of 54 supernova
(16 nearby ones and 38 at high redshifts). In order to determine
the cosmological parameters $\Omega_{\rm{m}}$ and $\beta$, we use a
$\chi^{2}$ minimization for a range of $\Omega_{\rm{m}}$ and
$\beta$ spanning the interval [0,1] in steps of 0.02
\begin{equation}
\chi^{2} =
\sum_{i=1}^{12}{\frac{\left[m(z_{i}, {\cal M}, \Omega_{\rm{m}},
\beta) - m_{oi}\right]^{2}}{\sigma_{i}^{2}}},
\end{equation}
where $m(z_{i}, {\cal M}, \Omega_{\rm{m}}, \beta)$ is given by Eq.
(6) and $m_{oi}$ is the observed values of the effective magnitude
with errors $\sigma_{i}$ of the $i$th measurement in the sample.
The ``zero point magnitude" ${\cal M}$ is considered as a
``nuisance" parameter so that we marginalize over it. 68$\%$ and
95$\%$ confidence regions are defined by the conventional
two-parameters $\chi^{2}$ levels 2.30 and 6.17, respectively.

\begin{figure}
\vspace{.2in}
\centerline{\psfig{figure=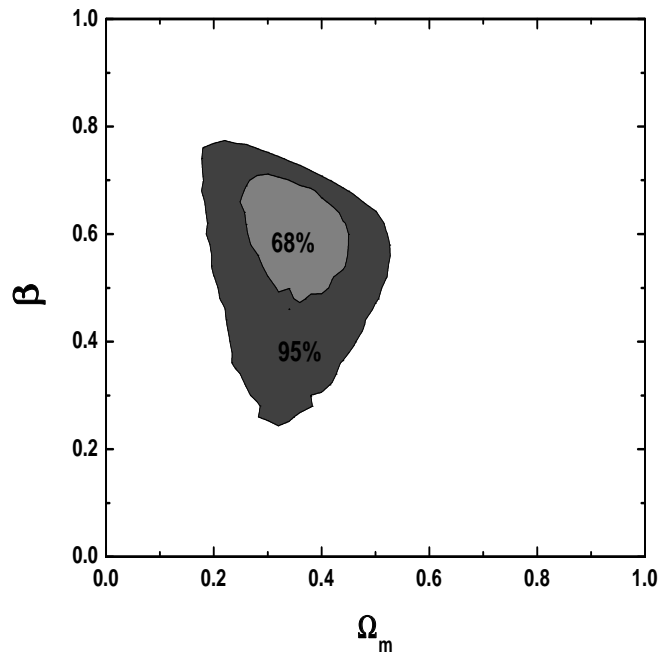,width=3.5truein,height=3.5truein}
\hskip 0.1in}
\caption{Confidence regions (68$\%$ and 95$\%$) in the $\Omega_{\rm{m}} - \beta$ plane provided by the
SNe Ia data from Perlmutter {\it et al.} \cite{perlmutter} by
assuming a Gaussian prior on the matter density parameter
$\Omega_{\rm{m}} = 0.35 \pm 0.07$.}
\end{figure}

In Fig. 1 we display the Hubble diagram for 16 low-redshift SNe Ia
and 36 high-redshift SNe Ia from SCP\cite{perlmutter} for
$\Omega_{\rm{m}} = 0.3$ and several values of $\beta$. For the sake
of comparion the standard open case $\beta = 0$ is also plotted. As
has been shown recently \cite{mesa}, a low-density decelerated
model cannot be ruled out by SNe Ia data alone, although such a
model is strongly deprived in the light of the recent CMB data.
Figure 2 shows contours of constant likelihood (95$\%$ and 68$\%$)
in the plane $\Omega_{\rm{m}} - \beta$. Note that the allowed range
for both $\Omega_{\rm{m}}$ and $\beta$ is reasonably large, showing
the impossibility of placing restrictive bounds on these scenarios
from the current SNe Ia data without an additional constraint on
the matter density parameter. The best fit model occurs for
$\Omega_{\rm{m}} \simeq 1.0$ and $\beta \simeq 0.52$ with $\chi^{2}
= 62.83$ and 52 degrees of freedom. In Fig. 3 we show confidence regions in
the $\Omega_{\rm{m}} - \beta$ plane by assuming a Gaussian prior on
the matter density parameter, i.e., $\Omega_{\rm{m}} = 0.35 \pm 0.07$ ($95\%$ c.l.).
Such a value is derived by combining the ratio of baryons to the total mass in clusters determined from X-ray
and Sunyaev-Zeldovich measurements with the latest estimates of the
baryon density $\Omega_{\rm{b}} = (0.020 \pm 0.002)h^{-2}$
\cite{burles} and the final value of the Hubble parameter obtained
by the {\it HST} key Project $H_o = 72 \pm 8$
${\rm{km.s^{-1}.Mpc^{-1}}}$ \cite{free}. In this case the best fit
model is strongly modified when compared with the previous
analysis. It occurs for $\Omega_{\rm{m}} = 0.34^{+0.14}_{-0.12}$
and $\beta = 0.62^{+0.12}_{-0.24}$ (95$\%$ c.l.) with $\chi^{2} =
63.44$. These results agree with the limits on the decaying
parameter $\beta$ from measurements of the angular size of
high-redshift radio sources. This latter analysis indicates $\beta \simeq 0.6$ while the
current estimates for the age of the Universe provides $\beta \geq 0.25$ \cite{jailson}.

\begin{figure}
\vspace{.2in}
\centerline{\psfig{figure=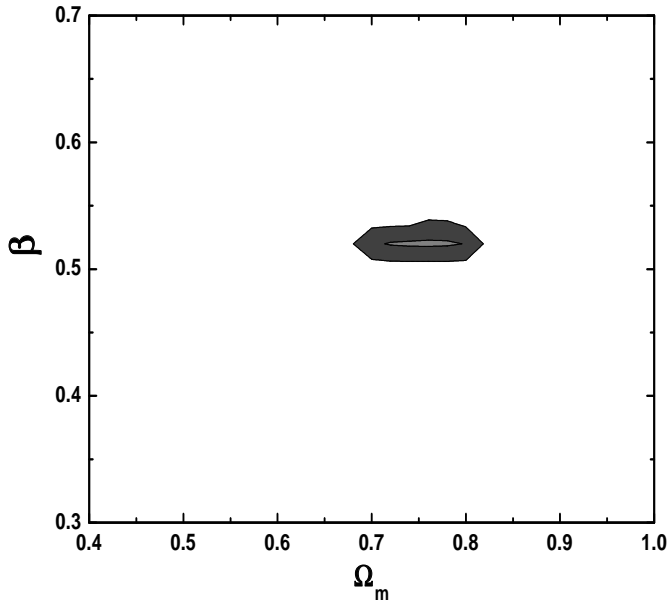,width=3.5truein,height=3.3truein}
\hskip 0.1in}
\caption{Simulation of the confidence limits from a SNAP-type data set generated
by assuming that we live in a
flat $\Lambda$CDM universe with $\Omega_{\rm{m}} = 0.28$ and
$\Omega_{\Lambda} = 0.72$. The dark shaded region corresponds to
95$\%$ c.l. while the brighter shaded region corresponds to 68$\%$
c.l.}
\end{figure}

\section{Constraints from future SNe data}

Let us now discuss the constraints on deflationary $\Lambda(t)$
cosmologies that may be expected from future SNe Ia data. Such data
may be provided by the proposed SNAP satelite, a two-meter space
telescope dedicated to SNe Ia observations on a wide range of
redshifts \cite{snap}. In this Section we investigate the limits on
the decaying rate $\beta$ based on one year of SNAP data. To this
end, we follow Goliath {\it et al.} \cite{goli} and assume 2000
supernova in the redshift interval $z \in [0,1.2]$ and an
additional of 100 supernova at higher redshifts, $z \in [1.2,1.7]$.
Each interval has been binned with $\Delta z = 0.05$. The
statistical error in magnitude, including the estimated measurement
error of the distance modulus and the dispersion in the distance
modulus due to dispersion in galaxy redshift, is assumed to be
$\Delta m = 0.15$ mag. The supernova data set was generated by
assuming that we live in a flat $\Lambda$CDM universe with
$\Omega_{\rm{m}} = 0.28$ and $\Omega_{\Lambda} = 0.72$.

Figure 4 shows the corresponding confidence regions for the
$\Omega_{\rm{m}} - \beta$ plane from this simulated supernova data
et. In comparison with Figs. 2 and 3, we see that the allowed
parameter space is strongly restricted by the SNAP data. In
particular, the statistical uncertainty on the $\beta$ parameter
reduces from $^{+0.12}_{-0.24}$ (by assuming a prior knowledge on
$\Omega_{\rm{m}}$) to $\sim{0.015}$. This result shows that future
SNe Ia data will provide much tighter constraints on deflationary
cosmologies or, possibly, on any kind of decaying $\Lambda(t)$
cosmologies, than do the current observations. Naturally, only with
a more general analysis, a joint investigation involving different
classes of cosmological tests, it will be possible to delimit the
$\Omega_{\rm{m}} - \beta$ plane more precisely, as well as to test
more properly the consistency of these scenarios. Such an analysis
will appear in a forthcoming communication \cite{alc}.

\section{Conclusion}

The results of observational cosmology in the last years have
opened up an unprecented opportunity to test the veracity of a
number of cosmological scenarios. The most remarkable finding among
these results cames from distance meaurements of SNe Ia at
intermediary redshifts that suggest that the expansion of the
Universe is speeding up, not slowing down. In this work, we
investigated the observational constraints on a particular class of
deflationary cosmologies provided by the current and future SNe Ia
data. We showed that the supernova data alone cannot place
restrictive constraints on the decaying parameter $\beta$ unless a
prior knowledge of the matter density parameter is introduced. In
this case, by assuming the gaussian prior $\Omega_{\rm{m}} = 0.35
\pm 0.07$ we found $\beta = 0.62^{+0.12}_{-0.24}$ at 95$\%$ c.l.
Such a result is in agreement with recent estimates of the $\beta$
parameter from measurements of the angular size of high-redshift
radio sources and age estimates of globular clusters
\cite{jailson}. By simulating one year of SNAP data we also found
that very restrictive limits will be placed on such cosmologies,
with error estimates on the decaying $\beta$ parameter of the order
of $\pm 0.02$.

\begin{acknowledgments}
The authors are grateful to Prof. C. J. Hogan and J. V. Cunha for helpful
discussions. JSA is
supported by the Conselho Nacional de Desenvolvimento
Cient\'{\i}fico e Tecnol\'{o}gico (CNPq) and CNPq
(62.0053/01-1-PADCT III/Milenio).
\end{acknowledgments}


\end{document}